\documentclass[final,5p,times,twocolumn]{elsarticle}

\usepackage{amssymb}
\usepackage{hyperref}
\usepackage{epsfig}
\usepackage{graphicx}
\graphicspath{ {./images/} }
\usepackage{subcaption}

\usepackage[svgnames]{xcolor}
  \definecolor{diffstart}{named}{Grey}
  \definecolor{diffincl}{named}{Green}
  \definecolor{diffrem}{named}{OrangeRed}

\usepackage[frozencache,cachedir=.]{minted}
\usepackage{sourcecodepro}
\setminted{fontsize=\footnotesize}

\newcommand{\customlink}[2]{%
  \href{#2}{#1}\footnote{\url{#2}}%
}

\journal{SoftwareX}

\begin{document}
\renewcommand{\labelenumii}{\arabic{enumi}.\arabic{enumii}}

\begin{frontmatter}
 


\title{OpenRAND: A Performance Portable, Reproducible Random Number Generation Library for Parallel Computations}


\author[label1]{Shihab Shahriar Khan\corref{cor1}}
\ead{khanmd@msu.edu}

\author[label2,label3]{Bryce Palmer}
\author[label4]{Christopher Edelmaier}
\author[label1]{Hasan Metin Aktulga}

\address[label1]{Dept. of Computer Science, Michigan State University, East Lansing, MI 48824}
\address[label2]{Dept. of Mechanical Engineering, Michigan State University, East Lansing, MI 48824}
\address[label3]{Dept. of Computational Mathematics, Science and Engineering, Michigan State University, East Lansing, MI 48824}
\address[label4]{Center for Computational Biology, Flatiron Institute, New York, NY 10010}
\cortext[cor1]{Corresponding author}

\begin{abstract}
\textit{We introduce OpenRAND, a C++17 library aimed at facilitating reproducible scientific research through the generation of statistically robust and yet replicable random numbers. OpenRAND accommodates single and multi-threaded applications on CPUs and GPUs and offers a simplified, user-friendly API that complies with the C++ standard's random number engine interface. It is portable: it functions seamlessly as a lightweight, header-only library, making it adaptable to a wide spectrum of software and hardware platforms. It is statistically robust: a suite of built-in tests ensures no pattern exists within single or multiple streams. Despite the simplicity and portability, it is remarkably performant\textemdash matching and sometimes even outperforming native libraries by a significant margin. Our tests, including a Brownian walk simulation, affirm its reproducibility and highlight its computational efficiency, outperforming CUDA's cuRAND by up to 1.8 times.}

\end{abstract}


\end{frontmatter}


\section*{Metadata}
\label{}

\begin{table*}[!ht]
\begin{tabular}{|l|p{6.5cm}|p{6.5cm}|}
\hline
\textbf{Nr.} & \textbf{Code metadata description} & \textbf{Please fill in this column} \\
\hline
C1 & Current code version & V0.9 \\
\hline
C2 & Permanent link to code/repository used for this code version & \nolinkurl{https://github.com/msu-sparta/OpenRAND} \\
\hline
C3  & Permanent link to Reproducible Capsule & \nolinkurl{https://codeocean.com/capsule/0144704/tree}\\
\hline
C4 & Legal Code License   & MIT License. \\
\hline
C5 & Code versioning system used & Git \\
\hline
C6 & Software code languages, tools, and services used & C++ 17 \\
\hline
C7 & Compilation requirements, operating environments \& dependencies & A compiler with C++17 support, optionally CMake\\
\hline
C8 & If available Link to developer documentation/manual & https://msu-sparta.github.io/OpenRAND \\
\hline
C9 & Support email for questions & khanmd@msu.edu \\
\hline
\end{tabular}
\caption{Code metadata}
\label{codeMetadata} 
\end{table*}

\section{Background}
Generating random numbers in a reproducible manner is pivotal for ensuring the reliability and validity of scientific research outcomes, especially in domains fundamentally reliant on random number generation, such as stochastic simulations, machine learning, and computer graphics. This reproducibility permits the exact replication of simulations, facilitating meaningful comparisons devoid of unnecessary variance and ensuring that any disparities arising are solely attributable to external factors, such as discrepancies in floating-point arithmetic ordering. More explicitly, it allows simulations utilizing identical random seeds to be compared directly, eliminating reliance on statistical averages and significantly simplifying the processes of debugging and regression testing, as has been demonstrated in past works \cite{phillips2011pseudo, netpyne}.

In single-threaded environments, the reproducibility of Pseudo Random Number Generators (RNGs) is straightforward since the same initial state (i.e., \textit{seed}) results in an identical sequence of random numbers (i.e., \textit{stream}). Reproducible random number generation for multi-threaded and multi-processed applications, on the other hand, is nontrivial due to the unpredictability of execution orders and the potential for race conditions. Strategies such as utilizing a single RNG instance with synchronization or distributing pre-allocated random numbers to various threads are fraught with scalability issues \cite{cbrng_review_lecuyer2021,phillips2011pseudo}. Instead, contemporary research has predominantly focused on the creation of multiple independent streams using distinct RNG instances per thread. One approach is to split a single stream into multiple equally sized streams. For example, one could split a single stream with period $2^{64}$ to $2^{32}$ independent streams, each with period $2^{32}$. This method, however, requires that the generator have a long period and efficient jump-ahead capability. Moreover, statistical independence between these streams is a concern \cite{de1988parallelization, l2017random}. Another commonly employed technique is to pre-generate a set of (pseudo)random seeds for each stream and employ them as starting points for generating multiple streams. Although the probability of a direct seed collision, resulting in identical streams, can be minimized for RNGs with sufficiently large periods, there remains the potential for statistical correlation among streams, particularly when the bit representations of two seeds are closely related \cite{pcg14}. For an extensive treatment of this subject, please refer to \cite{l2017random}.

The integration of Graphical Processing Units (GPUs) into computational workflows, vital for High-Performance Computing (HPC) workloads, introduces a distinct set of challenges as we shift from CPU to GPU, from dozens or, at most, hundreds of threads to potentially millions within a single node. Memory considerations serves as a prime illustration. In the GPU environment, threads access only a limited amount of high-speed private memory; hence, optimizing local memory usage is essential for maintaining performance. For instance, the default random engine in GNU's libstdc++, Mersenne Twister \cite{mt19937}, requires approximately 624 32-bit words for internal state, exceeding by more than double the maximum number of 32-bit registers permitted per thread in CUDA. Further, the absence of certain instructions in GPUs, common in most  CPUs, leads to distinct CPU vs. GPU performance characteristics for many generators \cite{r123}, prompting the need for specialized parallelization strategies. For example, the GPU-adapted Mersenne Twister, MTGP \cite{mtgp}, requires a block of threads to share a single state.


\section{Motivation and significance}
In HPC, the efficiency and reproducibility of pseudo-random number generation are paramount. Libraries like cuRAND and rocRAND excel at generating reproducible pseudo-random sequences in parallel environments but face challenges in portability, limiting their adaptability to new architectures. Conversely, while RandomCL \cite{randomcl} and clRNG \cite{clrng} (though no longer actively developed) successfully address hardware portability, they are limited to the OpenCL software framework. Two challenges arise across these platforms: seeding intricacies and the demands of state management. For seeding, neither RandomCL nor clRNG allow arbitrary seeds; instead, they generate seeds for a predetermined number of threads using a host-based sequential generator, subsequently transferring them to the device's global memory. cuRAND and rocRAND allow arbitrary seeds to produce distinct random streams but necessitate complicated state initialization procedures. Moreover, these platforms obligate developers to consistently manage memory states throughout the lifecycle of a processing element or thread, adding computational overhead. Overall, while these libraries offer potent tools, they introduce multifaceted challenges, emphasizing the need for comprehensive and streamlined solutions.

Counter-based generators (CBRNGs) offer a promising alternative to the challenges presented by traditional pseudo-random generators. Historically, these generators adapted cryptographic algorithms for HPC applications, albeit at reduced cryptographic strength \cite{chacha}, \cite{tea10}, \cite{r123}. A pivotal advancement in this domain was the integration of the counter concept into these generators \cite{chacha, r123}, mirroring the counter mode observed in stream ciphers. This innovation allowed a unique specification of a stream within a kernel for a given seed. Consequently, the earlier design pattern of one stream per processing element evolved to one stream per processing element per kernel, the utility of which has been demonstrated by HOOMD-blue \cite{anderson2020hoomd}. CBRNGs also champion the avalanche property, wherein a minimal bit alteration in the seed or counter cascades into a significantly altered, statistically independent new stream. This property gives developers the flexibility to use any unique values for the seed or counter, often harnessing internal application variables for parallel streams. Complementing these benefits, these generators feature notably compact states; for instance, OpenRAND boasts a 96-bit state, comfortably fitting within the maximum number of 32-bit registers permitted per thread in CUDA.

Building on this CBRNG foundation, Random123 is a pivotal library in the realm of counter-based random number generation \cite{r123}. It made significant strides by introducing three innovative generators and was at the forefront in terms of performing comprehensive statistical testing. Nevertheless, despite its contributions, Random123 exhibits notable drawbacks. The library does not incorporate some of the modern CBRNGs that have emerged in recent research \cite{neves2012fast, widynski2020squares}. Its API skews heavily toward lower-level implementations, exposing developers to the algorithmic details underpinning its generators. Furthermore, its reliance on intrinsics for performance enhancements compromises portability and inflates the codebase. As the demands of this domain escalate, the need for a holistic, refined solution becomes evident.

Addressing this need, we have developed the OpenRAND library, which amalgamates diverse counter-based generators with complementary strengths under a unified, user-centric API. Previously, developers suffered in two ways: either through increased development overhead from redundant API and easy-to-make bugs or through the execution of unnecessary instructions, leading to performance degradation. Our proposed solution, OpenRAND bridges this gap by focusing exclusively on a single family of counter-based generators. This concentrated approach facilitates the development of reproducible parallel code, allowing developers to circumvent the complications of API boilerplate and avoid unnecessary performance penalties. The library is characterized by its portability; it is a header-only library rigorously tested across various CPU and GPU devices and is compatible with multiple software platforms. Integration is streamlined, allowing incorporation into new or pre-existing projects with ease through CMake or by simply copying and pasting the requisite header files. It also touts a lightweight footprint, with its core header files comprising just 470 source lines of code at the time of publication. OpenRAND has undergone extensive statistical testing to ensure its reliability and robustness. As we will show, despite our emphasis on simplicity and user-friendly interface, OpenRAND is as fast as, and at times superior to, native CPU and GPU libraries.

\section{Software description}
\subsection{API Design}
OpenRAND is structured around a core set of counter-based random number generators complemented by a suite of examples, benchmarks, and tests. Complying with the C++17's random engine interface, generators in OpenRAND are compatible with standard library functions, including generating random samples across different distributions. Each generator is created via a constructor that requires two arguments: a 64-bit seed and a 32-bit counter \footnote{Except for one generator, Squares, which currently accepts 32-bit seeds}. When combined, the seed and counter produce a unique stream with a period of $2^{32}$. The seed is meant to uniquely identify each logical thread or processing element in the program\cite{l2017random}, whereas the counter can then be used to create multiple streams \textit{per seed} as needed. In practical scenarios, a processing element could be a particle in particle dynamics simulation or a pixel index in a ray tracing application.

Currently, OpenRAND supports a variety of counter-based generators, including Philox \cite{r123}, Threefry \cite{r123}, Squares \cite{widynski2020squares}, and Tyche \cite{neves2012fast}. They offer high-quality streams in compact sizes, efficient construction/destruction of RNG objects, and accept arbitrary seeds. The combination of these properties means, quite remarkably, that many stochastic applications can sidestep the need for \textit{any} random state maintenance. In practical terms and in contrast to Nvidia's cuRAND library, developers can often forgo the hassles and performance degradation related to storing states in global memory, launching a separate kernel on GPUs to initialize the states, or, in the case of array of structures data structures, the overhead of loading and saving states inside each kernel for every thread. An illustrative example of these advantages is shown in Section \ref{example}.

\section{Illustrative Examples} \label{example}
To demonstrate OpenRAND's advantages, we employ the Brownian Dynamics macro-benchmark from \cite{phillips2011pseudo}, re-implementing in CUDA across three RNG libraries. Fig. \ref{code:openrand_ie} highlights OpenRAND, Fig. \ref{code:curand_ie} cuRAND, and Fig. \ref{code:r123_ie} Random123. The compactness of OpenRAND's API is immediately evident, with just two lines for generator initialization and random number computation\textemdash over 14 fewer lines than the competing libraries. Unlike cuRAND, both OpenRAND and Random123 use CBRNGs, allowing unique particle IDs as seeds, guaranteeing a unique stream per particle regardless of thread count. This design circumvents the state maintenance associated with cuRAND, eliminating the need for memory allocation, state initialization, and continuous memory operations within each kernel thread. Nevertheless, random number generation with Random123 requires excessive boilerplate for initialization and random sampling, burdening developers with extra coding demands, and amplifying the risk of inadvertently introducing bugs. In light of these assessments, OpenRAND emerges as a potent blend of simplicity, efficiency, and adaptability, underscoring its viability as a prime choice for random number generation within HPC applications.

\begin{figure}[!ht]
\begin{minted}[xleftmargin=3pt,linenos,numbersep=3pt]{c++}
typedef openrand::Philox RNG;

__global__ void apply_forces(Particle *particles, 
                             int counter){
  int i = blockIdx.x * blockDim.x + threadIdx.x;
  if (i >= N)
    return;

  Particle p = particles[i];
  // Apply drag force
  p.vx -= GAMMA / mass * p.vx * dt;
  p.vy -= GAMMA / mass * p.vy * dt;

  // Apply random motion
  RNG local_rand_state(p.pid, counter);
  rnd::double2 r = local_rand_state.draw_double2(); 
  p.vx += (r.x  * 2.0 - 1.0) * sqrt_dt;
  p.vy += (r.y  * 2.0 - 1.0) * sqrt_dt;
  particles[i] = p;
}

int main(){
  // Initialize particles
  init_particles<<<nblocks, nthreads>>>(particles,
                                    /*counter*/ 0);

  // Simulation loop
  int iter = 0;
  while (iter++ < STEPS) {
    apply_forces<<<nblocks, nthreads>>>(particles,
                                             iter);
    ...
  }
}
\end{minted}
\vspace{-0.5cm}
\caption{Illustrative example of OpenRAND’s API}\label{code:openrand_ie}
\end{figure}

\begin{figure}[!ht]
\begin{minted}[xleftmargin=3pt,linenos,numbersep=3pt]{c++}
typedef curandStatePhilox4_32_10_t RNG;

__global__ void rand_init(RNG *rand_state) {
  int i = threadIdx.x + blockIdx.x * blockDim.x;
  if (i >= N) 
    return;
  curand_init(1984, i, 0, &rand_state[i]);
}


template <typename RNG>
__global__ void apply_forces(Particle *particles, 
                             RNG* rand_state){
  int i = blockIdx.x * blockDim.x + threadIdx.x;
  if (i >= N)
    return;

  Particle p = particles[i];
  // Apply drag force
  p.vx -= GAMMA / mass * p.vx * dt;
  p.vy -= GAMMA / mass * p.vy * dt;

  // Apply random motion
  RNG local_rand_state = rand_state[i];
  double2 r = curand_uniform2_double(
                    &local_rand_state); 
  p.vx += (r.x  * 2.0 - 1.0) * sqrt_dt;
  p.vy += (r.y  * 2.0 - 1.0) * sqrt_dt;
  rand_state[i] = local_rand_state;
  particles[i] = p;
}

int main(){
  // Random number generator setup
  RNG *d_rand_states;
  cudaMalloc((void **)&d_rand_states, N*sizeof(RNG));

  // Initialize random number generators
  rand_init<<<nblocks, nthreads>>>(d_rand_states);

  // Initialize particles
  init_particles<<<nblocks, nthreads>>>(particles, 
                                    d_rand_states);

  // Simulation loop
  int iter = 0;
  while (iter++ < STEPS) {
    apply_forces<<<nblocks, nthreads>>>(particles, 
                                    d_rand_states);
    ...
  }
}
\end{minted}
\vspace{-0.5cm}
\caption{Illustrative example of cuRAND's API}\label{code:curand_ie}
\end{figure}

\begin{figure}[!ht]
\begin{minted}[xleftmargin=3pt,linenos,numbersep=3pt]{c++}
typedef r123::Philox4x32 RNG;

__global__ void apply_forces(Particle *particles, 
                             int counter){
  int i = blockIdx.x * blockDim.x + threadIdx.x;
  if (i >= N)
    return;

  Particle p = particles[i];
  // Apply drag force
  p.vx -= GAMMA / mass * p.vx * dt;
  p.vy -= GAMMA / mass * p.vy * dt;

  // Apply random motion
  RNG rng;
  RNG::ctr_type c={{}};
  RNG::ukey_type uk={{}};
  uk[0] = p.pid;
  RNG::key_type k=uk;

  c[0] = counter; 
  c[1] = 0;
  RNG::ctr_type r = rng(c, k);

  uint64_t xu = (static_cast<uint64_t>(r[0]) << 32) 
      | static_cast<uint64_t>(r[1]);
  uint64_t yu = (static_cast<uint64_t>(r[2]) << 32) 
      | static_cast<uint64_t>(r[3]);
  auto x = r123::u01<double, uint64_t>(xu);
  auto y = r123::u01<double, uint64_t>(yu);

  p.vx += (x  * 2.0 - 1.0) * sqrt_dt;
  p.vy += (y  * 2.0 - 1.0) * sqrt_dt;
  particles[i] = p;
}

int main(){
  // Initialize particles
  init_particles<<<nblocks, nthreads>>>(particles, 
                                   /* counter*/ 0);

  // Simulation loop
  int iter = 0;
  while (iter++ < STEPS) {
    apply_forces<<<nblocks, nthreads>>>(particles, 
                                            iter);
    ...
  }
}
\end{minted}
\vspace{-0.5cm}
\caption{Illustrative example of Random123's API}\label{code:r123_ie}
\end{figure}

\section{Empirical Results}
\subsection{Performance Benchmarks}
To assess OpenRAND's performance, we designed two benchmark tests focusing on micro and macro performance metrics across CPU and GPU platforms, respectively.

\begin{figure}[!ht]
    \centering
    \begin{subfigure}{0.49\textwidth}
        \centering
        \includegraphics[width=\linewidth, height=2in, keepaspectratio]{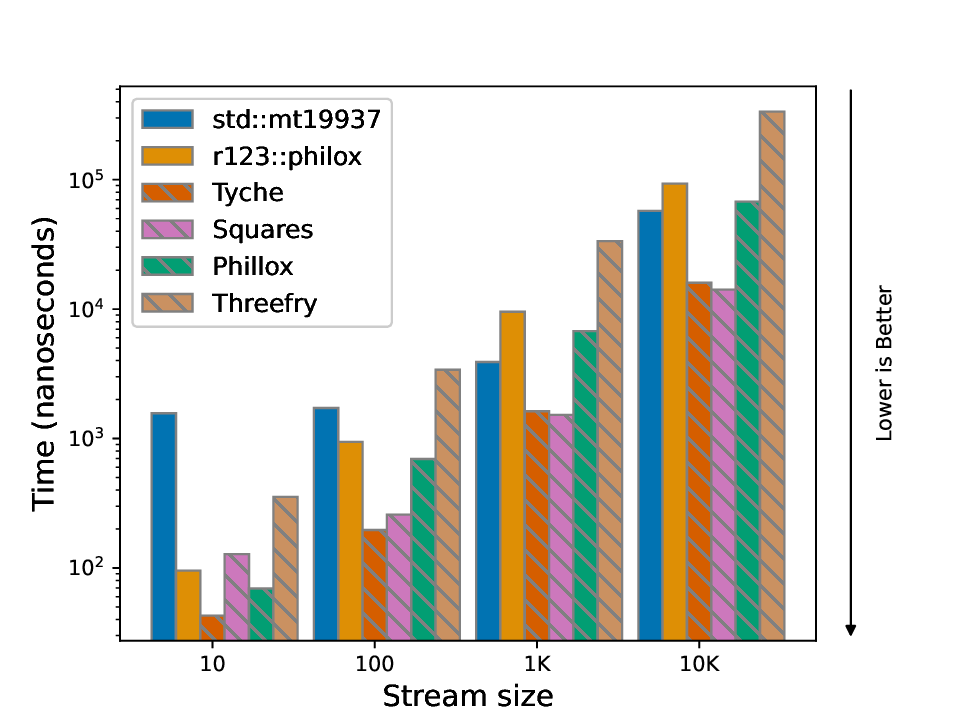}
        \caption{Time taken by OpenRAND generators versus baselines (std::mt19937 and r123::philox) to produce specified stream lengths on the host.}
        \label{cpu_fig}
    \end{subfigure}
    \hfill  
    \begin{subfigure}{0.49\textwidth}
        \centering
        \includegraphics[width=\linewidth, height=2in, keepaspectratio]{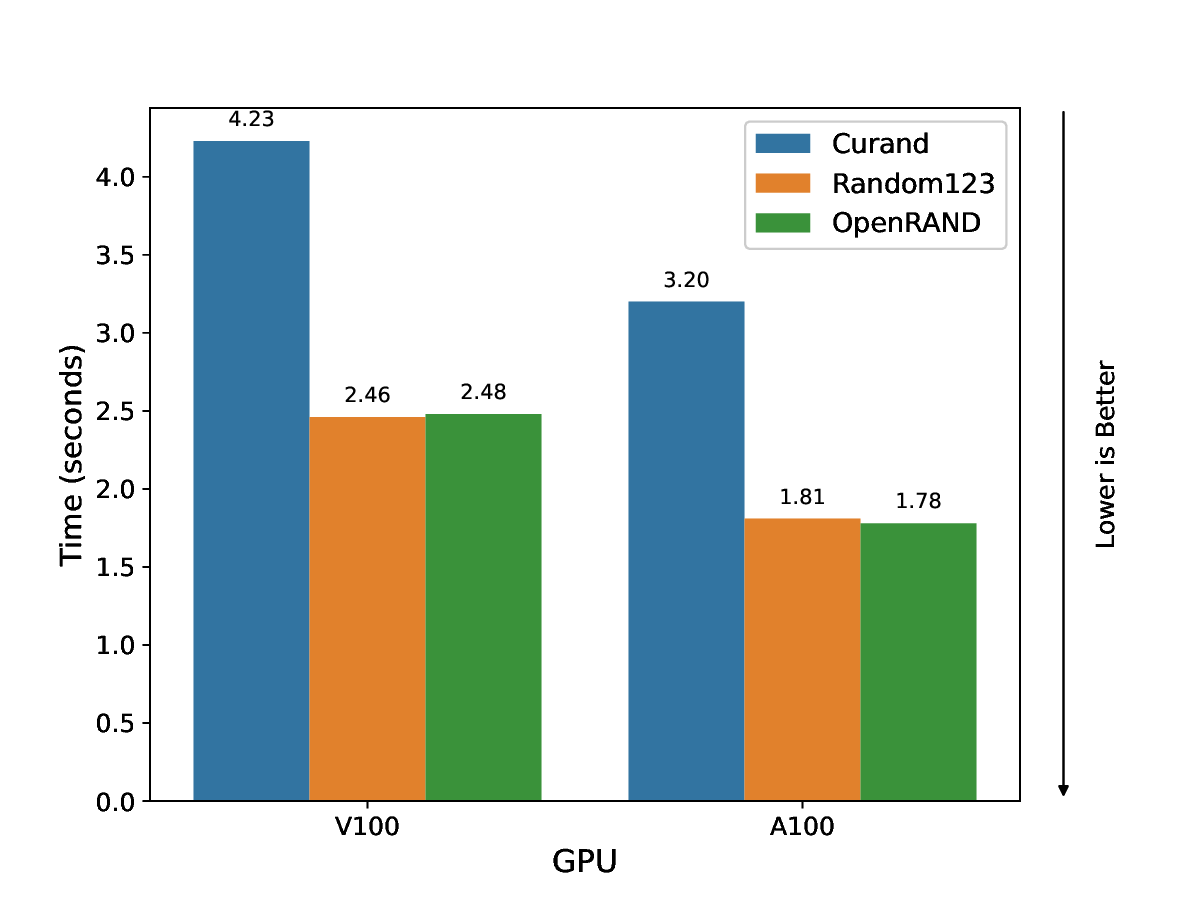}
        \caption{Wall time for various libraries executing the Brownian Dynamics benchmark on different GPUs, using the Philox generator \cite{r123} in each library.}
        \label{gpu_fig}
    \end{subfigure}
    \caption{Performance of OpenRAND on host and device respectively.}
\end{figure}

For our micro-benchmark, we implemented a single-threaded program tailored for CPU performance and tested it on an Intel(R) Xeon(R) Platinum 8260. This benchmark gauges the raw random number generation speed for streams of varying sizes across all generators. We employed Google benchmark for this evaluation, setting our performance standards against the widely recognized GNU libstdc++'s default random engine, mt19937 \cite{mt19937}, given its ubiquity in many applications. As the data in Fig. \ref{cpu_fig} reveals, OpenRAND's generators consistently outpace mt19937 for smaller streams, a realm often encountered in parallel programs. While this strong disparity in performance for small streams can be attributed to mt19937's intricate initialization routine, the performance advantage of OpenRAND over mt19937 is sustained even in longer streams for both the Tyche \cite{neves2012fast} and Squares \cite{widynski2020squares} generators.

Transitioning to GPU performance, we employ the previously discussed macro-benchmark, a 2D Brownian dynamics simulation in CUDA. This simulation involved one million independent particles diffusing according to a Brownian random walk. Particles were monitored over 10,000 steps, with the particles influenced by both a velocity-proportional drag force and a random uniform motion. To maintain consistency, pseudo-random number generation for all libraries used their respective Philox generators (For details, refer to \customlink{code}{https://github.com/Shihab-Shahriar/brownian-dynamics}). This benchmark was executed on two Nvidia GPUs: a Tesla V100 PCIe with a theoretical 14.13 TFLOPS and 900GB/s bandwidth and an A100 SXM with 19.5 TFLOPS and 2039GB/s bandwidth.

As seen in Fig. \ref{gpu_fig}, OpenRAND outperformed cuRAND by 1.8x, while saving \textasciitilde64 MB of GPU memory per million particles, and performed on par with Random123. Given the simplistic nature of the kernels used in the program, where random number generation dominates computational cost, such a performance margin between OpenRAND and cuRAND was unanticipated. Of course, we do not expect this margin of improvement to hold for real-world kernels where the computational cost of random number generation is less prominent. Nevertheless, this comparison between cuRAND, a native library specifically optimized for these platforms, and Random123, a library that utilized intrinsic instructions to achieve performance enhancement, offers confidence that OpenRAND's platform-independent code and simplified API do not compromise its performance.

\subsection{Statistical Evaluation}
To ensure the quality of our random number generation, OpenRAND exclusively incorporates generators with rigorous empirical validations and long-standing use. Even with this foundation, maintaining statistical integrity requires careful implementation, as subtle bugs can compromise randomness. As such, we perform rigorous quality assurance: every generator within OpenRAND was subjected to statistical testing using the popular frameworks TestU01 \cite{testu01} and PractRand \cite{practrand}. These tools offer a suite of complementary statistical tests designed to identify any underlying patterns or irregularities in random streams of data. An example of these tests is the Birthday Spacing test from TestU01 \cite{testu01}, which contrasts empirical results against known analytical solutions to detect potential discrepancies.

We initiated our testing process by evaluating individual data streams, probing them to their theoretical limit of $2^{32}$ integers using PractRand across a comprehensive range of keys and counters. While TestU01 and PractRand are geared towards single-stream assessments, we recognized the importance of extending these tests to encompass parallel streams, reflecting their use in real-world parallel computations. To perform our parallel stream tests, we followed the procedure outlined in \cite{anderson2020hoomd}\textemdash we simulated a scenario with 16,000 particles, generating micro-streams comprising three random numbers for each particle. These individual micro-streams for each particle were first combined into a single concatenated stream. This unified stream was then lengthened over successive iterations to examine correlations across the entire system. 

All generators were successfully tested for at least 1TB of data using PractRand and TestU01's comprehensive BigCrush battery of tests. It is worth elaborating on the BigCrush results. During repeated trials with multiple global seeds, certain outputs (one or two out of 106 tests) occasionally emerged as suspicious. However, this is not unique to OpenRAND; the authors of \customlink{cuRAND}{https://docs.nvidia.com/cuda/curand/testing.html} noted similar failures. For an exhaustive breakdown of our statistical results, we direct readers to our \customlink{documentation}{https://msu-sparta.github.io/OpenRAND/md_statistical_results.html}. To the best of our knowledge, this is the first time Tyche \cite{neves2012fast} and Squares \cite{widynski2020squares} generators have undergone correlation tests for parallel streams.

\section{Impact}
Random number generation plays a fundamental role in the efficiency and reliability of larger software systems across fields such as stochastic simulations, machine learning, and computer graphics. Ideally, there would exist a good off-the-shelf solution that could be used in a variety of contexts, including multi-threaded/multi-process applications, without introducing excessive boilerplate code, unnecessary complexity, or restrictions on applicable architecture. The existing software landscape is, however, fraught with challenges. Some good options expose low-level algorithmic and implementation details (e.g., Random123, cuRAND), leading to increased complexity; others are intrinsically bound to specific hardware (like cuRAND) or software platforms (such as rocRAND, OneAPI MKL). Several once-popular, platform-agnostic alternatives are now abandonware (clRNG, RandomCL), and even universal options, like the C++ Standard library, prove ill-suited for GPGPU programs. This landscape has led many esteemed open-source platforms to either layer atop a low-level library, like HOOMD-Blue's use of Random123, or to write custom random generators\textemdash as seen in Tensorflow, Pytorch, VTK, Jax, Alpaka, Kokkos, and others\textemdash sometimes without the benefit of thorough statistical validation.

OpenRAND aims to be that off-the-shelf solution. Prioritizing speed, reproducibility, parallelism, and portability, it offers these features through an accessible and streamlined API. For those seeking ease of integration, it can be seamlessly added to projects via CMake or by copying the required header files. In our benchmarks, OpenRAND outperforms cuRAND by a factor of 1.8 and performs on par with Random123. Despite this competitive performance, OpenRAND differentiates itself by offering a cleaner, more intuitive API, a lightweight code base, and an absence of machine-specific code. Validated across CPU and GPU platforms (with evaluations in standard C++, CUDA, HIP, and Kokkos), OpenRAND is uniquely positioned to elevate various projects by simplifying development, boosting performance, and enhancing portability while ensuring statistical validity and reproducibility for single and parallel streams. This commitment to quality is evident in its design\textemdash with comprehensive statistical quality tests incorporated within its continuous integration pipeline\textemdash ensuring a consistent standard as OpenRAND evolves within the open-source community.

\section{Conclusion}
To summarize, while the realm of random number generation presents numerous software options riddled with challenges\textemdash from exposing intricate algorithmic details to being tightly bound to specific architectures\textemdash OpenRAND emerges as an off-the-shelf solution. Its focus on a single family of counter-based generators simplifies the development of reproducible parallel code, avoiding the complications and redundancies faced with other libraries. By sidestepping the pitfalls of boilerplate APIs, hardware limitations, and abandoned alternatives, OpenRAND presents an efficient, streamlined, and platform-agnostic solution that seamlessly integrates into projects, reducing unnecessary complexities, overheads, and hardware restrictions. OpenRAND accelerates the development process while maintaining statistical robustness and reproducibility. Given its features and performance benchmarks, OpenRAND has the potential to significantly aid developers in various scientific fields, ensuring that random number generation remains both reliable and efficient.

\section*{Declaration of Competing Interest}
The authors declare that they have no known competing financial interests or personal relationships that could have appeared to influence the work reported in this paper.

\section*{Acknowledgements}
This material is based upon work supported by the National Science Foundation Office of Advanced Cyberinfrastructure under Grant 2007181 and used resources provided by Michigan State University's High-Performance Computing Center. 

\vspace{5pt}

\textbf{Declaration of generative AI and AI-assisted technologies in the writing process}

During the preparation of this work the authors used ChatGPT only in order to improve language and readability. The authors reviewed and edited the final content and take full responsibility for the content of the publication.



 






\end{document}